\documentclass{aadebug}

\usepackage{latexsym}
\usepackage{amssymb}
\usepackage{code}

\newcommand{\bi}{\begin{array}[t]{@{}l@{}}}
\newcommand{\ei}{\end{array}}
\newcommand{\ba}{\begin{array}}
\newcommand{\ea}{\end{array}}

\newcommand{\bp}{\begin{quote}\sf\begin{tabbing}}
\newcommand{\ep}{\end{tabbing}\end{quote}}

\newcommand{\arrow}{\rightarrow}

\newcommand{\proparrow}[0]{\Longrightarrow}
\newcommand{\comment}[1]{}

\newcommand{\tr}[1]{}

\newtheorem{ex}{Example}
\newenvironment{example}{\begin{ex}\rm}
		        {\end{ex}}

\newenvironment{ttprog}{\begin{trivlist}\small\item \tt
        \begin{tabbing}}{\end{tabbing}\end{trivlist}}

\begin{document}

\corr{0309027}{247}

\runningheads{Peter J.~Stuckey, et al.}
	     {The Chameleon Type Debugger (Tool Demonstration)}

\title{The Chameleon Type Debugger \\ (Tool Demonstration)}

\author{
Peter J.~Stuckey\addressnum{1}\comma\extranum{1},
Martin Sulzmann\addressnum{2}\comma\extranum{2}, \\
Jeremy Wazny\addressnum{1}\comma\extranum{3}
}

\address{1}{
Department of Computer Science and Software Engineering,
University of Melbourne, 3100, Australia
}

\address{2}{
School of Computing, 
National University of Singapore 
S16 Level 5, 3 Science Drive 2, 
Singapore 117543
}

\extra{1}{E-mail: pjs@cs.mu.oz.au}
\extra{2}{E-mail: sulzmann@comp.nus.edu.sg}
\extra{3}{E-mail: jeremyrw@cs.mu.oz.au}

\pdfinfo{
/Title (The Chameleon Type Debugger (Tool Demonstration))
/Author (Peter J.Stuckey, Martin Sulzmann, Jeremy Wazny)
}

\begin{abstract}
In this tool demonstration, we give an overview of the Chameleon type debugger.
The type debugger's primary use is to identify 
locations within a source program which are involved in a type error.
By further examining these (potentially) problematic program locations, users
gain a better understanding of their program and are able to work towards the
actual mistake which was the cause of the type error.
The debugger is interactive, allowing the user to provide additional
information to narrow down the search space. One of the novel aspects
of the debugger is the ability to explain erroneous-looking types. 
In the event that an unexpected type is inferred, the debugger can highlight
program locations which contributed to that result.
Furthermore, due to the flexible constraint-based foundation that the
debugger is built upon, it can naturally handle advanced type system features
such as Haskell's type classes and functional dependencies. 
\end{abstract}



\section{Introduction} \label{sec:intro}

Reporting meaningful type error messages is a notoriously difficult problem
for strongly typed languages.
Messages are often obscure and don't always indicate the actual source of the 
problem.
Beginners in particular, often have difficulty locating type errors.
In the case of languages with a powerful type system such as 
Haskell~\cite{haskell98}, even
experienced programmers can not always easily identify the source of the error.

Chameleon is a Haskell-style language.
The Chameleon system~\cite{chameleon} provides an interactive type debugging environment.
In the event of a type error or unexpected result, the system 
automatically highlights the (possibly)
erroneous program locations involved. The type debugger is also able to deal
with Haskell-style overloading~\cite{wadler-blott:ad-hoc} and
its various extensions~\cite{JonesESOP2000,jones-jones-meijer:typeclasses}.
We refer to~\cite{debugg-tr}, for a detailed account 
of the methods and techniques employed
and a broader comparison to related work.
For the purpose of this paper, we restrict our attention to a few examples
to highlight the main aspects of the Chameleon type debugger.

\section{The Type Debugger}

\subsection{Overview}

We present an example of a simple debugging session, which illustrates the
Chameleon type debugger in action.


\begin{example}
This program is supposed to print a graph of a mathematical function on the
terminal. The {\tt plot} function takes as arguments: a) the function to plot, 
b) the width of each terminal character, c) the height of each character, and 
d) the vertical distance (in characters) of the origin from the top of the 
screen.

\begin{ttprog}
plot \=  f dx dy oy = \\
    \> let \= fxs  = getYs f dx \\
    \> \>   	ys   = map (\verb+\+y-> fromIntegral (y-oy)*dy) [maxY,maxY-1..minY] \\
    \> \>	rows = map (doRow fxs) ys \\
    \> in  unlines rows \\
    \> where \\
\>\>	doRow [] r     = "" \\
\>\>	doRow (y:ys) r = \= (if y < r \verb+&&+ y > (r-dy) then '*' \\
\>\>				   \> else ' ') : doRow r ys \\
\>\>	getYs \= f dx = [ f ((centre x * dx)) | x <- [minX..maxX] ] \\
\>\>\>	    where centre = (+) .5 \\
\\
minX = 0  \\
maxX = 79 \\
minY = 0 \\
maxY = 19
\end{ttprog}

Curious to see some functions plotted, we load this program in Hugs and are met
with the following:

\begin{ttprog}
ERROR "plot.hs":13 - Type error in function binding
\\ *** Does not match \=: [a] -> [[a]] -> String \kill
\\ *** Term           \>: doRow
\\ *** Type           \>: [[a]] -> [a] -> [Char]
\\ *** Does not match \>: [a] -> [[a]] -> String
\\ *** Because        \>: unification would give infinite type
\end{ttprog}

We launch the Chameleon type debugger, and ask for the type
of {\tt doRow}. Note we use the syntax {\tt ;} to refer to
local definitions of {\tt plot}. Note also that in Haskell systems 
typically it is impossible to examine the types of local variables.
\begin{ttprog}
plot.hs> :type plot;doRow \\
type error - contributing locations \\
\underline{doRow} (\underline{y:}ys) \underline{r} = (if \underline{y < r} \&\& y > r - dy then '*' else ' ') : \underline{doRow r} ys
\end{ttprog}
We see that {\tt doRow}'s first argument is the pattern {\tt (y:ys)} whereas 
in the body of the definition, it is first applied to {\tt r}. 
This is a problem since the function {\tt <} equates the 
types of {\tt r} and {\tt y}, and yet by reversing the order of {\tt doRow}'s
arguments we are also equating the types of {\tt r} and {\tt ys}.
This is the source of the "infinite type" error message we saw earlier.
To resolve this problem we simply reorder the arguments in the recursive call.
The new clause is:
\begin{ttprog}
doRow (y:ys) r = (\= if y < r \&\& y > (r-dy) then '*' \\ 
		  \> else ' ') : doRow ys r
\end{ttprog}

Having fixed this bug, we return to Hugs, and attempt to load the corrected 
program. Hugs reports the following:

\begin{ttprog}
ERROR "plot.hs":19 - Illegal Haskell 98 class constraint in inferred type
\\ *** Expression\=: getYs
\\ *** Type      \>: \=(Num (a -> a), Num a, Num (Integer -> a)) => ((a->a)->b)
-> \\		 \>  \>(a->a) -> [b]
\end{ttprog}

We need to find out where these strange constraints are coming from in the 
defintion of {\tt getYs}.
The {\em explain} command allows us to ask why a variable or expression has a
type of a particular form. Note that we can write {\tt \_} to stand for types
we are not interested in.

\begin{ttprog}
plot.hs> :explain (plot;getYs) ((Num (Integer -> \_)) => \_) \\
getYs f dx = [ f ((\underline{centre} \underline{x} * dx)) | \underline{x} \underline{<-} \underline{[}minX\underline{..}maxX\underline{]} ]
\end{ttprog}

The problem appears to be the interaction between the {\tt centre} function and
the elements from the list being generated. The list must be responsible for
the presence of the {\tt Integer} type. 
We continue by scrutinising {\tt centre}, first examining its type
and then asking why it is functional.
\begin{ttprog}
plot.hs> :type plot;getYs;centre \\
(Num(b),Num(a -> b)) => a -> b -> b \\
plot.hs> :explain (plot;getYs;centre) \_ -> \_ -> \_ \\
\underline{centre} = \underline{(+)} \underline{.} 5
\end{ttprog}

The source of the error suddenly becomes clear. 
It was our intention that the expression {\tt .5} be a floating point
number.
In fact, the {\tt .} is parsed as a separate identifier - the function
composition operator.
The fix is to rewrite this as a floating point value that Haskell will
understand as such.
The corrected definition of {\tt centre} is:
\begin{ttprog}
centre = (+) 0.5
\end{ttprog}

Relieved that these two bugs are fixed, we return to Hugs to plot some curves,
and are faced with the following.
\begin{ttprog}
ERROR "plot.hs":8 - Instance of Fractional Integer required for definition of
plot
\end{ttprog}

We will begin with {\tt rows}, since we know that {\tt unlines}, a standard
Haskell function, can't have caused this problem.
The explain command will be useful in finding out which part of the program is
responsible for this constraint.
We work our way towards the problem from the top down.

\begin{ttprog}
plot.hs> :explain (plot;rows) ((Fractional Integer) => \_) \\
rows = map (doRow \underline{fxs}) ys \\
\\
plot.hs> :explain (plot;fxs) ((Fractional Integer) => \_) \\
fxs = \underline{getYs} f dx\\
\\
plot.hs> :explain (plot;getYs) ((Fractional Integer) => \_) \\
getYs f dx = [ f ((\underline{centre} \underline{x} * dx)) | \underline{x} \underline{<-} \underline{[minX..}maxX\underline{]} ] \\
\end{ttprog}

The problem seems to be the interaction between {\tt centre} and {\tt x}.

\begin{ttprog}
plot.hs> :type plot;getYs;centre \\
Fractional a => a \\
\\
plot.hs> :type plot;getYs;x \\
Integer
\end{ttprog}

Indeed this is the source of the mistake. The {\tt centre} function expects
{\tt Floating} values, whereas the elements generated from the list are 
{\tt Integer}s.
We will convert these {\tt Integer}s to the appropriate numeric type with the 
standard Haskell function {\tt fromIntegral}.
The corrected definition of {\tt getYs} follows.
\begin{ttprog}
getYs \= f dx = [ f ((centre (fromIntegral x) * dx)) | x <- [minX..maxX] ] \\
      \> where centre = (+) 0.5 
\end{ttprog}

With this last correction made, we reload Hugs, and are relieved to find that
our program is now accepted.

\end{example}

The following subsections broadly outline the type debugger's key features.
See Appendix~\ref{apx:summary} for a brief summary of all the debugger's
commands.

\subsection{Type Error Explanation}

The Chameleon type debugger primarily serves to direct users toward the
source of type errors within their programs. Thus, it would mostly be invoked
when a type error is already known to be present within a program.

Most traditional type checking/inference algorithms work by traversing a
program's abstract syntax tree, gathering type information as they proceed.
As soon as some contradictory type information is encountered, they terminate
with an error message. 
This message usually indicates that the location of the error is the site
in the program that was last visited.
In fact, typically we are more interested in uncovering all of the sites which
contribute to the type error - since any of those could be the actual source of
the error.

The Chameleon type debugger reports errors by highlighting all source locations
which contribute to an error.
The system underlines a minimal set of program locations which lead to a type
error. There may be multiple such sets - for reasons of efficiency, only one
set is found.
In Section~\ref{sec:most-likely} we show that despite this, the debugger can 
also underline the program locations that are common to all possible
error reports.

\begin{example}
Consider the following simple program.
\begin{ttprog}
f g x y = (g (if x then x else y), g "abc")
\end{ttprog}

Hugs reports:
\begin{ttprog}
ERROR "realprog.ch":1 - Type error in application
\\ *** Expression     : g (if x then x else y)
\\ *** Term           : if x then x else y
\\ *** Type           : Bool
\\ *** Does not match : [Char]
\end{ttprog}

This indicates that the type $\mathit{Bool}$ was inferred for the if-then-else 
expression, whereas a $\mathit{[Char]}$ was expected, and required. 
Why that is the case is left unknown to the programmer.

In Chameleon we can ask the system for the inferred type 
of some function $f$ using the command {\tt :type $f$}.
If the function has no type (contains a type error), then
an explanation in terms of locations that contribute to the error.
Using the Chameleon debugger, on the example above:
\begin{ttprog}
realprog.ch> :type f \\
f \underline{g} \underline{x} y = (\underline{g} (\underline{if x then x else} y), \underline{g} \underline{"abc"})
\end{ttprog}
The underlined locations are those which are implicated in the error.
There is no attempt to identify any specific location as the "true" source of 
the error.
No claim is made that the if-then-else expression must have type 
$\mathit{[Char]}$.
The source of all the conflicting types is highlighted and it is
up to the user to decide where the actual error is.
It may be that for this 
particular program, the mistake was in applying {\tt g} 
to a $\mathit{[Char]}$---a possibility that the Hugs message ignores.

\end{example}

The obvious drawback of this approach to error reporting is its 
dependence on the user to properly interpret the effect of each location
involved.
Indeed, it may not always be clear why some location contributes to the type
error.

Of course, the user is free to follow up his initial query with additional 
commands to further uncover the types within the program - 
but this may be tedious to do
repeatedly since it's the user's responsibility to make sense of all those
individual queries.
Future work will be directed toward improving the style of type error reports.

\subsection{Local/Global Explanation}

Put simply, the Chameleon type debugger works by generating constraints on the
types of variables in a program.
Type inference is then a matter of solving those constraints.

The more constraints that are involved, the slower this process may be.
In general, the number of constraints present is proportional 
to the size of the
call graph in a definition.
To reduce the amount of overhead, we provide two distinct reporting modes:
{\em local} and {\em global}.

In the local mode, we restrict error/type explanations to locations
within a single definition. 
The consequence of this is that we are then free to simplify the constraints
which arises from outside of this definition.

\begin{example}\label{ex:stack}
Consider the following simple program which is supposed to model a stack 
data structure using a list.
\begin{ttprog}
idStack stk = pop (push undefined stk) \\
push top stk  = (top:stk) \\ 
pop (top,stk) = stk \\
empty = []
\end{ttprog}
There's obviously a problem here.

Upon launching the debugger in local mode (the default), the following is
reported:
\begin{ttprog}
type error - contributing locations: \\
idStack stk = \underline{pop} (\underline{push} undefined stk) \\
\end{ttprog}
We then examine both of the culpable functions.
\begin{ttprog}
stack.hs> :type push \\
a -> [a] -> [a] \\ 
stack.hs> :type pop \\
(a,b) -> b 
\end{ttprog}
The error is revealed. In the definition of {\tt idStack}, {\tt push} returns a
list, where {\tt pop} expects a pair.

\end{example}

In the global explanation mode, we place no restrictions on the locations which
can be highlighted.
Since we no longer restrict ourselves to any particular locations, we can no
longer simplify away any of the constraints involved.

\begin{example}
We return to the previous example, but this time we use the global explanation
mode.

\begin{ttprog}
stack.hs> :set global \\
stack.hs> :type idStack \\
type error - contributing locations: \\
idStack stk = \underline{pop} (\underline{push} undefined stk) \\
push top stk  = (top\underline{:}stk) \\ 
pop \underline{(}top\underline{,}stk\underline{)} = stk 
\end{ttprog}

In this mode, all program locations contributing to the error are revealed
immediately. In this case, the conflict between the list and pair is
immediately revealed.
\end{example}

\subsection{Type Explanation}

It's possible to write a function which, though well-typed, does not have the
type that was intended.
A novel feature of our debugger is its ability to explain such unexpected 
types. 
The command {\tt :explain $(e)$ $(D \Rightarrow t)$} asks
the system to underline a set of locations which force expression $e$
to have type of the form $D \Rightarrow t$.
As is the case for type errors, explanations consist of references to the
responsible source locations.

Type explanations are particularly useful for explaining errors involving
malformed type class constraints.
A typical type error message would merely complain about the existence of such 
a constraint, without providing any information to further investigate the
problem.

\begin{example}

The following program scales all elements of a list, such that the biggest
element becomes 1.
It contains a type class error.
\begin{ttprog}
normalize xs = scale biggest xs \\
scale x ns = map (/x) ns \\
biggest \= (x:xs) = max x xs \\
	\> where \= max x [] = x \\
	\>	 \> max x (y:ys) \= | y > x     = max y ys \\
	\>	 \>		 \> | otherwise = max x ys
\end{ttprog}

Hugs reports the following:
\begin{ttprog} 
ERROR "normalize.hs":1 - Illegal Haskell 98 class constraint in inferred type
\\ *** Expression : normalize
\\ *** Type       : (Fractional ([a] -> a), Ord a) => [[a] -> a] -> [[a] -> a]
\end{ttprog}

The problem is that, in Haskell, type class constraints must be of form $C~a$,
where $a$ is a type variable.
The inferred constraint, $\mathit{Fractional} ([a] \arrow a)$, obviously violates this
condition.

To solve this problem, we'll need to know why $\mathit{Fractional}$'s argument is a
function.
Since this is a fairly compact program, we will use the debugger in 
global mode.
We proceed as follows.

\begin{ttprog}
normalize.hs> :set global \\
normalize.hs> :explain (normalize) ((Fractional ([\_] -> \_)) => \_) \\
normalize xs = \underline{scale} \underline{biggest} xs \\
\underline{scale} \underline{x} ns = map \underline{(/x)} ns \\
\underline{biggest} \= (x\underline{:}xs) = max x xs \\
	\> where \= max x [] = x \\
	\>	 \> max x (y:ys) \= | y > x     = max y ys \\
	\>	 \>		 \> | otherwise = max x ys

\end{ttprog}

It should be clear that the $Fractional$ constraint arises from the use of
{\tt /} in the definition of {\tt scale}.
We also see from the above exactly where that the list type comes from the
pattern in {\tt biggest}.
This indicates that the use of {\tt biggest} hasn't been fully applied in 
{\tt normalize}.

The following  corrected version of {\tt normalize} will work as intended.
\begin{ttprog}
normalize xs = scale (biggest xs) xs
\end{ttprog}

\end{example}

%

\subsection{Most-Likely Explanations} \label{sec:most-likely}

Since it is technically infeasible to present all explanations for a particular
type error or {\tt :explain} query, the debugger will, by default, only display
one.
However, in the case where the intersection of all explanations is non-empty,
it is also possible to cheaply identify the common program locations.
This information is useful since in most cases it would be reasonable to assume
that locations which are involved in more errors are more likely 
to be the real cause
of the problem. Indeed fixing a location involved in all errors will
immediately resolve them all.

If the debugger finds that there is a common subset of all errors, it will
automatically report it, along with a specific error.

\begin{example}
Consider the following  program. {\tt merge} takes two sorted lists as arguments
and combining all elements, returns a single sorted list.
\begin{ttprog}
merge [] ys = ys \\
merge xs [] = [xs] \\ 
merge (x:xs) (y:ys) \= | x < y     = [x] ++ merge xs (y:ys) \\
                    \> | otherwise = [y] ++ merge (x:xs) ys 
\end{ttprog}

This program contains a type error.
Loading it into the debugger, we get the following error report.
\begin{ttprog}
type error - conflicting locations: \\
merge [] ys = ys \\
\underline{merge} \underline{xs} [] = \underline{[xs]} \\ 
merge \underline{(x:xs)} (y:ys) 
		\= | x < y     = \underline{[x]} \underline{++} \underline{merge} \underline{xs} (y:ys) \\
                \> | otherwise = [y] ++ merge (x:xs) ys \\
\\
locations which appear in all conflicts: \\
merge [] ys = ys \\
\underline{merge} \underline{xs} [] = \underline{[xs]} \\ 
merge (x\underline{:}xs) (y:ys) \= | x < y     = [x] ++ merge xs (y:ys) \\
                    \> | otherwise = [y] ++ merge (x:xs) ys 
\end{ttprog}

The first display of the program above is a typical debugger error report.
A single error is reported and all responsible program locations are
highlighted.

The second program printout indicates those locations which appear in all
errors.
Note that these locations do not themselves constitute a type error.
An actual error occurs when these locations are considered alongside 
(either of) the recursive calls to {\tt merge} in the third clause.
Nevertheless, this provides us a view of the mostly likely source of error, and
makes clearer the mistake in this program.

\end{example}

\subsection{Type Inference for Arbitrary Locations}

Most existing interpreters/programming environments for Haskell, for example
Hugs\cite{hugs} and GHCi\cite{ghc}, allow users to ask for the type of
variables in their program.
These systems, however, restrict queries to variables bound at the top level of
their program.
It is not possible for users to inquire about the types of variables bound
within a {\tt let} or {\tt where} clause.
This can be frustrating since, in order to allows inference for those
definitions, programs need to be modified, lifting
sub-definitions to the top level.

Furthermore, these systems restrict type queries to programs which are
well-typed.
Obviously this is of little help when debugging a program, since it must
somehow be made typeable while maintaining the essence of the error - for
example by replacing expressions with calls to {\tt undefined}.

By contrast, the Chameleon type debugger allows users to infer types of
arbitrary variables and expressions within a program, regardless of whether it
is typeable.

\begin{example}
Consider the following program, where most of the work is being done in a
nested-definition.

\begin{ttprog}
reverse \= = rev [] \\
	\> where \= rev rs [] = rs \\
	\>	 \> rev rs (x:xs) = rev (x ++ rs) xs
\end{ttprog}

We run the debugger as follows:
\begin{ttprog}
reverse.hs> :type reverse \\
\mbox{[[a]] -> [a]}
\end{ttprog}
Since the problem is mostly due to the definition of {\tt rev}, we continue
by examining the type of {\tt rev}.
We can refer to nested bindings, like {\tt rev}, via their enclosing binding by
separating their names with {\tt ;}s. 
We can refer to a specific clause of a
definition by following it with a {\tt ;} and the number of the clause.
Otherwise, by default, the reference is to the entire declaration, including
all clauses.
\begin{ttprog}
reverse.hs> :type reverse;rev \\
\mbox{[a] -> [[a]] -> [a]} \\
reverse.hs> :type reverse;rev;1 \\
\mbox{a -> [b] -> a} \\
reverse.hs> :type reverse;rev;2 \\ 
\mbox{[a] -> [[a]] -> b}
\end{ttprog}
We see above that the second clause of {\tt rev} introduces the erroneous 
{\tt [[a]]} type.
Having narrowed our search to an easily digestible fraction of the original
code, we proceed as follows:
\begin{ttprog}
reverse.ch> :explain (reverse;rev;2) (\_ -> [[\_]] -> \_) \\
rev rs \underline{(x:xs)} = \underline{rev} (\underline{x} \underline{++} rs) \underline{xs}
\end{ttprog}
\end{example}

Certainly, a text-based interface places restrictions on how easily one may
refer to arbitrary program fragments, however this merely a limitation of the
current debugger implementation.
In principle, we feel that the ability to infer types anywhere within any 
program is of invaluable benefit when debugging type errors.


\subsection{Source-Based Debugger Interface}

Being interactive, the debugger's text-based interface provides users with
immediate feedback which they can use to iteratively work their way toward the
source of a type error.
Although flexible, such an interface can at times be slightly
stifling and a distraction from the source code being edited separately by the
programmer. 
As an alternative, we allow programmers to pose debugger queries directly in
their program's source. 

The command `{\tt :type} $e$', where $e$ is an expression, can 
be written directly in the program as $e${\tt::?}.
Also, {\tt :explain} $(e)$ $(D \Rightarrow t)$, where $e$ is an expression
and $D \Rightarrow t$ is a type scheme, 
can be expressed as $e${\tt::?}$t$ within the program itself.

As well as individual expressions, entire declarations can be queried by 
writing such a command at the same scope
(with a declaration name in place of an expression.)

The debugger responds to these embedded commands by collecting and processing
them in textual order. 
They do not otherwise affect the working of the compiler, and do not change the
meaning of programs they are attached to.

\begin{example}
Returning to Example~\ref{ex:stack}, we modify it by adding a {\tt type} query
to the expression in the body of {\tt idStack}, and attempt a re-compile.

\begin{ttprog}
idStack stk = (pop (push undefined stk)) ::? \\
push top stk  = (top:stk) \\ 
pop (top,stk) = stk \\
empty = []
\end{ttprog}

Compilation would still fail, but the Chameleon system would print the
following before stopping:
\begin{ttprog}
type error - contributing locations: \\
idStack stk = \underline{pop} (\underline{push} undefined stk) \\
\end{ttprog}

We continue by further modifying the program, adding additional queries, and
re-running the compiler.
In the following version we have attached {\tt type} queries to the definitions 
of {\tt push} and {\tt pop}.

\begin{ttprog}
idStack stk = pop (push undefined stk) \\
push ::?
push top stk  = (top:stk) \\ 
pop ::?
pop (top,stk) = stk \\
empty = []
\end{ttprog}

This time, the system would respond to our commands as follows.

\begin{ttprog}
push :: a -> [a] -> [a] \\ 
pop :: (a, b) -> b
\end{ttprog}

\end{example}

\section{Type Class Extensions}

A strength of our system is that it supports almost arbitrary type class 
extensions.
This is made possible through Chameleon's extensible type system~\cite{overloading}.

In Chameleon, one can specify type extensions
by providing additional constraint rules.
Such rules
take effect during
constraint solving (type inference.) 
These rules introduce new constraints, which can be used to enforce additional 
restrictions on types.
One use is to encode Haskell-style functional dependencies.

\begin{example}
Consider the following type class definition.

\begin{ttprog}
class \= Collect a b where \\
\> empty :: b \\
\> insert :: a -> b -> b \\
\> member :: a -> b -> Bool
\end{ttprog}

As defined this class is flawed.
The type of \texttt{empty :: Collect a b => b} is ambiguous, since
type variable {\tt a} appears only in the constraint component.
This poses a problem when running a particular use of {\tt empty}, since
it is not clear which instance it represents.

Functional dependencies allow us to overcome this problem, by relating the type
class variables in a specific way.
We are able to state that \texttt{b} functionally determines \texttt{a}. 
Though not part of Haskell 98, Hugs~\cite{hugs} and GHC~\cite{ghc} both 
support functional dependencies.
In either of these systems, the functional dependency can be stated as so:
\begin{ttprog}
class \= Collect a b | b -> a where ...
\end{ttprog}

The same functional dependency can be expressed in Chameleon via the rule:
\begin{ttprog}
rule Collect a b, Collect a' b $\proparrow$ a = a'
\end{ttprog}
This states that if we have to {\tt Collect} constraints with the same second
argument, then their first argument must also be the same.

Consider the following program which tries to check if a $\mathit{Float}$
is a member of a collection of $\mathit{Int}$s.

\begin{ttprog}
f g x y = \= if member (x::Float) (insert (1::Int) y) \\
	  \> then g x else y
\end{ttprog}

The constraints for $f$ imply $\mathit{Collect}~\mathit{Int}~t$ and 
$\mathit{Collect}~\mathit{Float}~t$.
This allows firing of the rule representing
the functional dependency adding the information that $Int = Float$,
causing a type error.

Note that our standard error reporting mechanism will still work in the
presence of arbitrary rules, and will report the following:

\begin{ttprog}
collect.ch> :type f \\
type error - contributing locations \\
f g x y = \= if \underline{member} (x::\underline{Float}) 
         (\underline{insert} (1::\underline{Int}) y) \\
\> then g x else y \\
rule(s) involved: Collect a b, Collect a' b ==> a = a'
\end{ttprog}

\end{example}

Rules can be written to enforce other arbitrary conditions.

\begin{example}
Consider
\begin{code}
f x y = x / y + x `div` y
\end{code}
The inferred type is
{\tt f :: (Integral a, Fractional a) => a -> a -> a}.
This is obviously not satisfiable, though it will not immediately cause an 
error.
To address this issue, we can impose the following rule.
\begin{code}
rule Integral a, Fractional a ==> False
\end{code}

\end{example}

This rule essentially states that we do not allow the same type, {\tt a}, to
both an instance of {\tt Integral} and {\tt Fractional}.
In terms of the constraints we generate during type inference, it means we
would add the always-false constraint whenever this condition arises - causing
the process to fail with a type error.

In the case of the above example, adding this rule will case the Chameleon 
debugger to report the following.

\begin{ttprog}
div.ch> :t f \\
type error - contributing locations: \\
f x y = x \underline{/} y + x \underline{`div`} y
\\ 
rule(s) involved: 
   Integral a, Fractional a ==> False
\end{ttprog}



\section{Related and Future Work}

In recent years, there has been an increasing awareness of the limitations of
traditional inference algorithms when it comes to type error reporting.
Various strategies have been proposed to improve the usefulness of type error
messages and aid the programmer in debugging type errors.
These include things like: alternative inference algorithms, error 
explanation systems, and interactive type explanation tools.

Error explanation systems allow the user to examine the process by which
specific types are inferred for program 
variables~\cite{duggan96explaining,beaven94explaining}.
Most work by recording the effect of the inference procedure at each step.
In this way a complete history of the process can be built up.
One unfortunate shortcoming of such systems is the typically large size of
of explanations.
This occurs since, although all steps are recorded, not all are of 
interest to the user when investigating a specific type.
Also, since such systems are layered on top of an existing inference
algorithm, they suffer from the same left-to-right bias when discovering
errors.

The Chameleon approach is to map the typing problem to a constraint
solving problem, and determine minimal systems of constraints (and
corresponding program locations) which cause errors or types.
Independently Haack and Wells~\cite{HW03} use a similar approach
to explain type errors, though their system does not apply to
Haskell style overloading which makes its usefulness limited
for Haskell debugging.

Interactive tools exist which assist the user to find type errors
manually~\cite{huch00typeview,chitil01compositional}.
These essentially allow one to uncover the type of any program location.
Through examination of the types of subexpressions, the user gradually works
towards the source of the error.
Identifiers with suspicious looking types are followed to their definition and
further examined.
This approach to debugging is similar to that currently supported by the 
Chameleon type debugger, but is generally much more limited.
None of these allow for quick and efficient identification of
conflicting program locations; the process is manually directed at every step.
Furthermore, none of them allow the user to immediately locate precisely those 
sites which contribute to some suspicious looking type.
Our type debugging tool is flexible enough to allow both a user-directed
examination of the program's types as well as provide useful, succinct advice
to direct the user's search.
In principal, a declarative debugging-style interface to our debugger could be
designed to emulate the aforementioned systems.

Helium\cite{helium} is a language and compiler designed to be similar to 
Haskell, with the aim of providing superior error messages for beginners.
Based on a database of common mistakes, the system can provide
additional hints to users, suggesting improvements that are likely to correct
the program.
By contrast, the Chameleon type debugger makes little effort in presenting
its diagnostic information, beyond highlighting the sites that are involved.
To remedy this, we hope to apply similar strategies to those found in Helium.
Our belief is that the two forms of error reporting strongly reinforce each 
other.
A well-focused error message provides a good starting point, while highlighted
program locations lead the user towards sites worth exploring .

Work remains to make the debugger easier to use, both in terms of its
interface and functionality.
We intend to expand the debugger's error reporting capability to provide better
type and error explanations, as well as improved error messages in general.

Currently our approach to inferring types in the presence of errors is
fairly limited. 
In future work we plan to extend the debugger so that it reports
a strongest possible type, given all errors it is involved in. 
This would further support our efforts to provide improved diagnostic
information to users, since we might be able to suggest minimal program fixes 
which resolve all type errors.

The current system is limited to finding a single type error in a definition.
This limitation is unfortunately necessary, since the task of finding all
errors is computationally expensive (although see~\cite{allmin}).
In the case of multiple errors, one is chosen arbitrarily.
The choice is a consequence of the implementation of the underlying algorithms.
It may be possible in future, to arrange things so that more likely errors are
discovered preferentially.
For example, we may be possible to discover errors amongst closely situated 
locations, before those involving quite distant program sites.


\section{Conclusion}

The problem of diagnosing type errors can be quite difficult, particularly for
beginning programmers.
Traditional inference algorithms give rise to error messages which can be
difficult to interpret at times, and occasionally even misleading.

The Chameleon type debugger is an interactive system which allows users to 
examine their programs' types in detail.
The debugger supports: inference of types for arbitrary locations, error
explanations outlining all locations involved, explanation of
suspicious-looking types, and more.
It is unique in that it supports all Haskell type system features.
The debugger's error reporting mechanism is still fairly basic, though
future work will focus on making the most of improving the system in this
regard.

\bibliography{main}

\appendix

\section{Summary of Debugger Commands}
\label{apx:summary}

The debugger supports the following commands.

\begin{description}
\item[\tt:type <expression>]
    Infers and displays the type of the given expression.
    In the event of an error, highlights contributing locations. 

\item[\tt:explain (<expression>) (<type>|<typescheme>)]
    Explains why the given expression has a type that is an instance of the 
    given
    type. Contributing locations are highlighted.

\item[\tt:declare (<reference>) (<type>|<typescheme>)]
    Declares that the referred-to definition has the nominated type. This
    declaration is internal to the debugger, and does not affect the program 
    source.

\item[\tt:print]
    Displays the loaded source program.

\item[\tt:set <flag>]
    Configures the debugger by setting internal options. \\ Available flags
    include:
    \begin{description}
    \item[\tt global] Selects global type/error explanations.
    \item[\tt local] Selects local type/error explanations.
    \item[\tt solver ?] Selects the version of the internal constraint solver to
	use. The available values are 0, 1 and 2. 
	Higher numbers lead to more thorough explanations, but involve 
	slower solvers.
    \end{description}
	
\item[\tt:help]
    Prints command help.

\item[\tt:quit]
    Quits the debugger.

\end{description}

\end{document}